\documentclass[superscriptaddress,aps,preprintnumbers,amsmath,amssymb,prd,nofootinbib,preprint]{revtex4-1}
\pdfoutput=1

\usepackage{graphicx}
\usepackage{epstopdf}
\usepackage{dcolumn}
\usepackage{bm}
\usepackage{hyperref}
\usepackage{color}
\usepackage{amsmath,amssymb}
\usepackage[sort&compress]{natbib}
\usepackage{float}

\usepackage{setspace}
\usepackage[hang]{footmisc}

\numberwithin{equation}{section}

\begin{document}


\def\a{\alpha}
\def\b{\beta}
\def\c{\varepsilon}
\def\d{\delta}
\def\e{\epsilon}
\def\f{\phi}
\def\g{\gamma}
\def\h{\theta}
\def\k{\kappa}
\def\l{\lambda}
\def\m{\mu}
\def\n{\nu}
\def\p{\psi}
\def\q{\partial}
\def\r{\rho}
\def\s{\sigma}
\def\t{\tau}
\def\u{\upsilon}
\def\v{\varphi}
\def\w{\omega}
\def\x{\xi}
\def\y{\eta}
\def\z{\zeta}
\def\D{\Delta}
\def\G{\Gamma}
\def\H{\Theta}
\def\L{\Lambda}
\def\F{\Phi}
\def\P{\Psi}
\def\S{\Sigma}

\def\o{\over}
\def\beq{\begin{align}}
\def\eeq{\end{align}}
\newcommand{\gsim}{ \mathop{}_{\textstyle \sim}^{\textstyle >} }
\newcommand{\lsim}{ \mathop{}_{\textstyle \sim}^{\textstyle <} }
\newcommand{\vev}[1]{ \left\langle {#1} \right\rangle }
\newcommand{\bra}[1]{ \langle {#1} | }
\newcommand{\ket}[1]{ | {#1} \rangle }
\newcommand{\EV}{ {\rm eV} }
\newcommand{\KEV}{ {\rm keV} }
\newcommand{\MEV}{ {\rm MeV} }
\newcommand{\GEV}{ {\rm GeV} }
\newcommand{\TEV}{ {\rm TeV} }
\newcommand{\1}{\mbox{1}\hspace{-0.25em}\mbox{l}}
\newcommand{\headline}[1]{\noindent{\bf #1}}
\def\diag{\mathop{\rm diag}\nolimits}
\def\Spin{\mathop{\rm Spin}}
\def\SO{\mathop{\rm SO}}
\def\O{\mathop{\rm O}}
\def\SU{\mathop{\rm SU}}
\def\U{\mathop{\rm U}}
\def\Sp{\mathop{\rm Sp}}
\def\SL{\mathop{\rm SL}}
\def\tr{\mathop{\rm tr}}
\def\mpl{M_{\rm Pl}}

\def\IJMP{Int.~J.~Mod.~Phys. }
\def\MPL{Mod.~Phys.~Lett. }
\def\NP{Nucl.~Phys. }
\def\PL{Phys.~Lett. }
\def\PR{Phys.~Rev. }
\def\PRL{Phys.~Rev.~Lett. }
\def\PTP{Prog.~Theor.~Phys. }
\def\ZP{Z.~Phys. }

\def\dd{\mathrm{d}}
\def\ff{\mathrm{f}}
\def\BH{{\rm BH}}
\def\inf{{\rm inf}}
\def\ev{{\rm evap}}
\def\eq{{\rm eq}}
\def\SM{{\rm sm}}
\def\Mpl{M_{\rm Pl}}
\def\GeV{{\rm GeV}}
\newcommand{\Red}[1]{\textcolor{red}{#1}}
\newcommand{\TL}[1]{\textcolor{blue}{\bf TL: #1}}

\title{
Axion Misalignment Driven to the Bottom
}
\preprint{LCTP-18-34}

\author{Raymond T. Co}
\affiliation{Leinweber Center for Theoretical Physics, Department of Physics, University of Michigan, Ann Arbor, Michigan 48109, USA}
\author{Eric Gonzalez}
\affiliation{Leinweber Center for Theoretical Physics, Department of Physics, University of Michigan, Ann Arbor, Michigan 48109, USA}
\author{Keisuke Harigaya}
\affiliation{School of Natural Sciences, Institute for Advanced Study, Princeton, NJ 08540, USA}

\begin{abstract}
Several theoretical motivations point to ultralight QCD axions with large decay constants $f_a \simeq \mathcal{O}(10^{16}\text{--}10^{17})$~GeV, to which experimental proposals are dedicated. This regime is known to face the problem of overproduction of axion dark matter from the misalignment mechanism unless the misalignment angle $\theta_{\rm mis}$ is as small as $\mathcal{O}(10^{-3}\text{--}10^{-4})$, which is generally considered a fine-tuning problem. We investigate a dynamical explanation for a small $\theta_{\rm mis}$. The axion mass arises from strong dynamics and may be sufficiently enhanced by early dynamics so as to overcome Hubble friction and drive the field value to the bottom of the potential long before the QCD phase transition. Together with an approximate CP symmetry in the theory, this minimum is very closely related to today's value and thus $\theta_{\rm mis}$ can automatically be well under unity. Owing to such efficient relaxation, the isocurvature perturbations are essentially damped. As an existence proof, using supersymmetric theories we illustrate that the Higgs coupling with the inflaton energy can successfully achieve this axion damping in a consistent inflationary cosmology.
\end{abstract}

\date{\today}

\maketitle

\section{Introduction}
\label{sec:intro}

In the Standard Model, a dimensionless charge conjugation and parity (CP) violating parameter $\theta$ is constrained to be less than $\mathcal{O}(10^{-10})$ by the experimental limit on the neutron electric dipole moment~\cite{Crewther:1979pi,Baker:2006ts}. However, there is no theoretical reason in the Standard Model why $\theta$ has to be exceedingly small. This fine-tuning problem is known as the strong CP problem~\cite{tHooft:1976rip}. An elegant solution was proposed in Refs.~\cite{Peccei:1977hh, Peccei:1977ur}, where $\theta$ is promoted to a field that dynamically relaxes to a CP-conserving minimum. This field, called the axion, can be understood as the pseudo Nambu-Goldstone boson arising from the spontaneous breaking of the global $U(1)$ Peccei-Quinn (PQ) symmetry~\cite{Weinberg:1977ma,Wilczek:1977pj}. The symmetry breaking scale $f_a$, also called the axion decay constant, is constrained by astrophysics to be larger than $\mathcal{O}(10^7-10^8)$ GeV~\cite{Ellis:1987pk,Raffelt:1987yt,Turner:1987by,Mayle:1987as,Raffelt:2006cw}. After the QCD phase transition, the axion acquires a mass from strong dynamics
\begin{equation}
m_a = 6 \, \mu\EV \left( \frac{10^{12} \ \GeV}{f_a}\right)
\end{equation}
and is thus constrained to be lighter than $\mathcal{O}(0.1)$ eV. Such a light axion is cosmologically stable and is hence a well-motivated dark matter (DM) candidate~\cite{Preskill:1982cy,Abbott:1982af,Dine:1982ah}. 

It is theoretically interesting to consider a decay constant above the grand unification scale $M_{\rm GUT} \simeq 2\times 10^{16}$ GeV, which is a typical prediction of string theory~\cite{Svrcek:2006yi}. It is also motivated from the field theory point of view in the supersymmetric Standard Model.
The breaking of the PQ symmetry may be of the same origin as the breaking of grand unification \cite{Wise:1981ry, Nilles:1981py, Hall:1995eq}. Actually, in four dimensional grand unified theories, if the $\mu$ term of the Higgs doublets is controlled by the PQ symmetry, as is the case with the DFSZ model~\cite{Dine:1981rt,Zhitnitsky:1980tq}, the symmetry breaking scale must be around the unification scale~\cite{Goodman:1985bw,Witten:2001bf,Harigaya:2015zea}. Many experimental efforts have since been devoted to this mass range of axions \cite{Budker:2013hfa, Kahn:2016aff, Ouellet:2018beu}. Nevertheless, ultralight QCD axions are subject to overproduction of dark matter from the misalignment mechanism. 

It is commonly assumed that the axion mass is lower than the Hubble scale $H_I$ during inflation so the field value is frozen due to Hubble friction (unless the number of e-folding is exceedingly large $N_e \sim (H_I/m_a)^2$ as considered in Refs.~\cite{Dimopoulos:1988pw,Graham:2018jyp, Guth:2018hsa}) and receives quantum fluctuations of order $H_I/2\pi$. Furthermore, if the PQ symmetry is broken before inflation, the axion field value is then homogenized throughout the observable universe and generically misaligned from the minimum of the potential generated by QCD effects at low temperatures. This potential energy of the axion then contributes to the dark matter abundance~\cite{Preskill:1982cy, Abbott:1982af, Dine:1982ah}
\begin{equation}
\label{eq:misalign}
\Omega_{\rm mis} h^2 \simeq 0.12~\theta_{\rm mis}^2 \left(\frac{f_a}{5\times10^{11}~\text{GeV}}\right)^{\frac{n+3}{n+2}}
\end{equation}
where $\theta_{\rm mis}$ is the misalignment angle and $n$ shows the temperature dependence of the axion mass $m_a \propto T^n$ at temperatures larger than the QCD confinement scale, whereas $n = 0$ is understood for $f_a \gtrsim 10^{17}$ GeV because the oscillations occur when the axion mass is already at the zero-temperature value. Here we have assumed a radiation-dominated era at the onset of axion oscillations. As a result of Eq.~(\ref{eq:misalign}), for $f_a \gg \mathcal{O}(10^{12})$ GeV, the universe is overclosed unless $\theta_{\rm mis}$ is sufficiently small (applicable only to the case of pre-inflationary PQ breaking). There exist late-time mechanisms such as dilution from entropy production by late decaying moduli fields \cite{Steinhardt:1983ia, Kawasaki:1995vt} or the axion superpartner \cite{Hashimoto:1998ua, Kawasaki:2011ym, Baer:2011eca, Bae:2014rfa, Co:2016vsi}, and depletion to other species \cite{Kitajima:2014xla, Agrawal:2017eqm, Ho:2018qur, Machado:2018nqk}. Meanwhile, it is crucial to explore natural explanations to a small $\theta_{\rm mis}$.

In this paper, we investigate a mechanism we dub Dynamical Axion Misalignment Production (DAMP), which exhibits the following two features 1) the axion field dynamically relaxes to the minimum of the potential in the early universe and 2) the model possesses a non-trivial prediction between the minima of the axion potential in the early and today's epochs. If the minima are approximately aligned, a small misalignment angle can dynamically arise without any fine-tuning, a scenario we refer to as DAMP$_{0}$. This damping effect automatically occurs when the assumption of $m_a \ll H_I$ is relaxed because the axion starts to oscillate during inflation and the amplitude is exponentially redshifted. We consider the case where the large axion mass originates from a large QCD scale during inflation. Such a scenario has been considered in Refs.~\cite{Dvali:1995ce,Banks:1996ea,Choi:1996fs}, but not all constraints, such as the effects from a large QCD scale on Higgs and other scalars, are fully evaluated. Hence, the examples presented in these studies may or may not be fully compatible with cosmological bounds.

As a proof of principle, we demonstrate that DAMP$_0$ can be realized in an extended Minimal Supersymmetric Standard Model (MSSM). The axion mass increases with the QCD confinement scale $\Lambda_{\rm QCD}$. By virtue of a negative Hubble induced mass via the coupling with the inflaton, the Higgs can be driven towards a large field value along the $D$-flat direction. The quark masses enhanced by a large Higgs vacuum expectation value (VEV) in turn cause strong dynamics to confine at a higher scale $\Lambda_{\rm QCD}'$ via the renormalization group (RG) running, thereby enhancing $m_a$. If only MSSM quark masses are raised, the resultant dynamical scale is not large enough to fulfill the first criterion of DAMP$_0$, which urges us to consider an extended MSSM. Earlier studies in Refs.~\cite{Dvali:1995ce,Banks:1996ea} use generic moduli fields to directly raise the QCD scale in an attempt to achieve DAMP$_0$ but do not carefully examine if the axion mass can be raised in a consistent way. Ref.~\cite{Choi:1996fs} investigates the consistency and concludes that the axion mass cannot be raised above the Hubble scale during inflation. We will clarify why our setup evades their claim. Later studies in Refs.~\cite{Jeong:2013xta,Choi:2015zra} introduce an extra $SU(3)_c$ charged particle and raise their masses by the large Higgs VEV. The purpose of the papers is to suppress the isocurvature perturbations and no attempts are made in fulfilling the second criterion for DAMP$_0$ to make predictions about the DM abundance.

The second criteria of DAMP$_0$ can be satisfied by an (approximate) CP symmetry of the theory ensuring that the minimum of the axion potential is nearly the same during inflation and in the vacuum~\cite{Banks:1996ea}. For example, the field values of any moduli which change the QCD $\theta$ term should remain unchanged. This can be understood by a CP symmetry which differentiates CP-odd moduli from CP-even moduli, and by assuming that the CP-odd moduli are always fixed at the enhanced symmetry points. Also, in the extended MSSM we introduce extra $SU(3)_c$ charged particles whose masses are large during inflation. The phase of the masses should be nearly aligned with that in the vacuum, which can be ensured by an approximate CP symmetry. We must introduce $\mathcal{O}(1)$ CP violation in the Yukawa couplings of the MSSM. We assume that the coupling among the source of CP violation, the moduli fields, and the extended sector is small, which may be understood by some symmetry or a geometrical separation in extra dimensions. Even if the CP phase is $\mathcal{O}(1)$ in the Yukawa couplings, perturbative quantum corrections to the moduli and the extended sectors are expected to be small, as the CP phase of the Yukawa couplings is physical only if three generations are simultaneously involved, suppressing the possible quantum corrections by multi-loop factors, the small Yukawa couplings, and generation mixings.%
\footnote{One may wonder that solving the strong CP problem by the CP symmetry without resorting to the QCD axion is more minimal. In fact, it is not easy to solve the strong CP problem while generating $\mathcal{O}(1)$ complex phases in the Yukawa couplings, as the complex phases readily correct the $\theta$ term. See Refs.~\cite{Nelson:1983zb,Barr:1984qx,Bento:1991ez,Hiller:2001qg} for models which work under several assumptions.}

A truly vanishing misalignment angle will imply the absence or a different origin of axion dark matter. Both of these are interesting possibilities, especially if DAMP$_0$ is applied to the fine-tuning problem in string axions. In this paper, we focus on QCD axion dark matter from the misalignment mechanism with $M_{\rm GUT} \lesssim f_a \lesssim \Mpl$, and thus a finite $10^{-4} \lesssim \theta_{\rm mis} \lesssim 10^{-3}$ is necessary. This implies that axion's minimum during inflation should nearly but not precisely coincide with today's value. One possibility is that the desirable amount of the CP-violating phase exists.
Another possibility is with $H_I \simeq m_a$ so that inflation ends exactly at the time when $\theta_{\rm mis}$ is relaxed to the desired value. 
We limit our consideration to the former and assume an approximate CP symmetry in the Higgs, inflaton, and extended sectors.%
\footnote{Suppressing the CP-violating phase is not sufficient in ensuring $\theta_{\rm mis} \ll 1$ because the parameters setting the axion minimum can be real but change signs during and after inflation, introducing a phase shift of $\arg(-1) = \pi$ in the potential. Equivalently, the axion minimum is converted to the hilltop, i.e.~$\theta_{\rm mis} \simeq \pi$. We explore this possibility called DAMP$_{\pi}$ in a separate publication \cite{Co:2018mho}.}
Interestingly, the CP violation in the MSSM around the TeV scale is currently constrained to $\mathcal{O}(10^{-3})$ by the electron dipole moment~\cite{Andreev:2018ayy, Cesarotti:2018huy} and might be detected in future experiments if CP violation of $\mathcal{O}(10^{-4} \text{--} 10^{-3})$ exists in the Higgs sector.

In Sec.~\ref{sec:misalign}, we review the misalignment mechanism for axion dark matter. In Sec.~\ref{sec:earlyrelax}, we illustrate in detail how the understanding of the misalignment angle can be dramatically different when the dynamics of the Higgs during inflation is taken into account. We summarize and discuss the conditions and implications of DAMP$_0$.

\section{Misalignment Mechanism}
\label{sec:misalign}

Since DAMP$_{0}$ operates at times well before the weak scale, it simply sets the initial condition for the standard misalignment mechanism~\cite{Preskill:1982cy, Abbott:1982af, Dine:1982ah}, which we will review in this section. When the temperature drops to near the QCD scale, the axion acquires a periodic potential energy through the color anomalies, with a mass given by
\begin{equation}
m_a ( T \ge \Lambda_{\rm QCD}) = 6 \, \EV \left( \frac{10^6 \ \GeV}{f_a}\right) \left( \frac{\Lambda_{\rm QCD}}{T} \right)^n ,
\end{equation}
where $n = 4$ for the SM is obtained by the dilute instanton gas approximation (see the lattice simulations in Refs.~\cite{Petreczky:2016vrs, Borsanyi:2016ksw, Burger:2018fvb, Bonati:2018blm, Gorghetto:2018ocs}, whose results indicate that $n$ ranges from 3.0 to 3.7 depending on the temperature). The equation of motion and the energy density read 
\begin{align}
\label{eq:EoM_a}
\ddot{\theta}_a + 3 H \dot{\theta}_a & = -m_a^2 \theta_a, \\
\label{eq:rho_a}
\rho_a & = \frac{1}{2} \left(m_a^2 \, \varphi^2 + \dot{\varphi}^2 \right) ,
\end{align}
where the axion field value $\varphi$ is interchangeable with the angle $\theta_a \equiv \varphi/f_a$.
Initially overdamped by the Hubble friction term in Eq.~(\ref{eq:EoM_a}), the axion mass increases through the QCD phase transition and the axion starts to oscillate coherently when $m_a \simeq 3 H$. After the onset of oscillations, the axion behaves as cold dark matter with the abundance given in Eq.~(\ref{eq:misalign}). Assuming the axion reproduces the observed DM abundance, a small $\theta_{\rm mis}$ leads to the prediction of large $f_a$
\begin{equation}
f_a \simeq 2 \times 10^{16} \ \GeV \left( \frac{2\times10^{-3}}{\theta_{\rm mis}} \right)^{\frac{2n+4}{n+3}} ,
\end{equation}
where $n = 0$ is understood for $f_a \gtrsim 10^{17}$ GeV as the axion mass reaches the zero-temperature value before the oscillation starts.
In what follows, we illustrate on how a small $\theta_{\rm mis}$ can naturally arise from dynamics via DAMP$_{0}$, as opposed to fine-tuning the initial condition.

\section{Early Relaxation during Inflation}
\label{sec:earlyrelax}

The axion field is generically assumed to be a constant during inflation as a result of the Hubble friction term in Eq.~(\ref{eq:EoM_a}) because one presumes that the axion mass is no larger than today's value. Nonetheless, this assumption holds only when the effects responsible for the axion mass remain invariant throughout the cosmological evolution. If the axion mass is larger than the Hubble parameter during inflation $H_I$, the axion begins its coherent oscillations, whose amplitude is damped exponentially. The axion mass may initially be enhanced by a smaller decay constant, a larger QCD scale, or a different origin of the axion mass. A smaller decay constant can occur when the PQ breaking dynamics evolves with time, whereas the axion may also receive extra mass contributions, e.g.~a large QCD confinement scale \cite{Dvali:1995ce,Banks:1996ea,Choi:1996fs, Jeong:2013xta,Choi:2015zra,Ipek:2018lhm}, explicit PQ breaking \cite{Higaki:2014ooa, Kawasaki:2015lea, Takahashi:2015waa, Kearney:2016vqw}, and magnetic monopoles \cite{Kawasaki:2015lpf,Nomura:2015xil}. 

We study the scenario where the QCD scale is enhanced by the inflationary dynamics of the Higgs or other moduli fields. QCD confines at the scale where strong dynamics becomes non-perturbative from the RG running. The number of active quark flavors affects the RG running and, in particular, $\Lambda_{\rm QCD}$ increases with the quark masses. Consequently, the quark masses that are raised during inflation, e.g.~by the Higgs VEV, increase the QCD scale and hence the axion mass. Other generic moduli fields can also directly affect the gauge coupling constant~\cite{Dvali:1995ce,Banks:1996ea,Choi:1996fs,Ipek:2018lhm} and thus the QCD scale as well. We first discuss the minimal setup of the MSSM and a large Higgs VEV before introducing additional particles.

The Higgs evolution during inflation crucially depends on its coupling with other fields. If a negative mass term is generated by the VEVs of other scalars and dominates over the Hubble scale, the Higgs field is driven to a large value where higher dimensional operators become important in stabilizing the Higgs. The MSSM provides a well-motivated framework for this realization. To be concrete, we assume the following K\"ahler potential 
\begin{align}
\label{eq:Kahler}
\Delta K =  \frac{|X|^2}{M^2} \left( |H_u|^2 + |H_d|^2+ \big( H_u H_d + c.c. \big)  - \frac{|H_u|^2 |H_d|^2}{M^2}  - \frac{|H_u|^4}{M^2}  - \frac{|H_d|^4}{M^2}  \right),
\end{align}
where $X$ is the chiral field whose $F$-term provides an inflaton energy and $M$ is the cutoff scale of the theory. Here and hereafter we assume a universal cutoff and drop $\mathcal{O}(1)$ coefficients. Through Eq.~(\ref{eq:Kahler}), the energy density of the inflaton $\rho_X = F_X F_X^* \simeq H_I^2 M_{\rm Pl}^2$ generates the Hubble induced mass as well as the higher dimensional operators in the Higgs potential
\begin{align}
\label{eq:HiggsL}
\Delta V =  c H_I^2 \left( - |H_u|^2 - |H_d|^2   - \big( H_u H_d + c.c. \big) 
 + \frac{|H_u|^2|H_d|^2}{M^2}  +  \frac{ |H_u|^4 }{M^2}  +  \frac{ |H_d|^4 }{M^2} \right), 
\end{align}
where $c = M_{\rm Pl}^2/M^2$. These additional Hubble induced terms in Eq.~(\ref{eq:HiggsL}) affect both the radial and angular directions of the Higgs fields. The negative Hubble induced mass, $-cH_I^2 (|H_u|^2 + |H_d|^2)$, drives the Higgs along the $D$-flat direction $|H_u| = |H_d| \equiv \phi$ towards large VEVs of order $M$, which are stabilized by the positive quartic terms. This enhances the axion mass via a larger dynamical scale from heavier quarks.
We note that the Higgs energy density is comparable to that of the inflaton and makes an $\mathcal{O}(1)$ change to the vacuum energy. Since the Higgs field value remains constant and the Higgs energy density follows that of the inflaton, this only changes the overall energy scale of the inflation and does not interfere with inflation dynamics. Conversely, if the sign of $cH_I^2 (|H_u|^2 + |H_d|^2)$ is positive instead, the conventional cosmology results because the Higgs VEVs remain small. Finally, the term $- cH_I^2 \big( H_u H_d + c.c. \big)$ fixes the relative phase of $H_u$ and $H_d$.%
\footnote{Ref.~\cite{Choi:1996fs} does not introduce this term and relies on the vacuum $B\mu$ term to fix the phase. They hence restrict their attention to the case where $H_I<m_{\rm SUSY}$. Furthermore, they assume that $M\sim \mpl$, and thus the Higgs does not take a large field value during inflation. The axion mass is suppressed in their setup and DAMP$_0$ cannot be realized.}
This term is not necessary if the vacuum $B\mu$ term is already larger than $H_I^2$. \pagebreak We assume an approximate CP symmetry so that this Higgs phase is nearly aligned with today's value,%
\footnote{We assume the sign of $cH_I^2 \big( H_u H_d + c.c. \big)$, if required to fix the relative phase, is the same as $B\mu \big( H_u H_d + c.c. \big)$ from soft supersymmetry breaking in the present universe. Rather, if the signs are opposite, the Higgs phase shifts by $\pi$, placing the axion at the hilltop instead. We consider this interesting possibility in Ref.~\cite{Co:2018mho}.} 
satisfying the second criterion of DAMP$_0$. Here $\sqrt{c} H_I$ is assumed to be larger than the supersymmetry (SUSY) soft breaking scale $m_{\rm SUSY}$ and hence we need 
\begin{equation}
\label{eq:HImin}
H_I \gtrsim 10 \ \GeV \left( \frac{m_{\rm SUSY}}{\TEV} \right) \left( \frac{M}{M_{\rm GUT}} \right),
\end{equation} 
where $M_{\rm GUT}= 2\times 10^{16}$ GeV. We may relax this condition if the $\mu$ term as well as the soft SUSY breaking terms are small at large field values of the Higgs or during inflation. The soft terms are actually smaller in gauge mediation since the large Higgs field value breaks the gauge symmetry.

The large Higgs VEV gives quarks very large masses during inflation. In the MSSM, the 1-loop renormalization group equation (RGE) is 
\begin{equation}
	\mu_r \frac{d}{d\mu_r} \frac{8 \pi^2}{g^2} = 3N-F,
\end{equation}
where $\mu_r$ is the renormalization scale, $N$ is the gauge group index, and $F$ is the number of active fermions in the theory.
In the MSSM with a large Higgs VEV $\phi_i$, assuming the gauge couplings are held fixed at the GUT scale,
and pretending that all quarks (including possible KSVZ quarks~\cite{Kim:1979if, Shifman:1979if}) are heavier than the dynamical scale, we find that the fiducial dynamical scale is raised to
\begin{equation}
\label{eq:LQCDhiggs}
\Lambda_\text{fid} = 10^7~{\rm GeV} \left( \frac{\phi_i}{10^{16}~{\rm GeV}} \right)^{2/3} \left( \frac{{\rm tan}\beta}{1} \right)^{1/3}.
\end{equation}
The fiducial dynamical scale coincides with the physical dynamical scale $\Lambda_\text{QCD}'$ if all quarks are actually heavier than $\Lambda_\text{fid}$.
Otherwise, they are related as
\begin{equation}
\label{eqn:L_fid}
\Lambda_{\rm fid} \equiv \Lambda_\text{QCD}' \prod_{m_q < \Lambda_\text{QCD}'} \left( \frac{m_q}{\Lambda_\text{QCD}'} \right)^{1/9}.
\end{equation}
When the physical dynamical scale is raised beyond the gluino mass $m_{\tilde{g}}$, the axion mass is suppressed by $m_{\tilde{g}}$ as
\begin{equation}
\label{eqn:ma}
m_a \simeq \frac{1}{4\pi} \frac{m_{\tilde{g}}^{1/2} \Lambda_{\rm fid} ^{3/2}}{f_a}  \simeq 10 \, \KEV \left( \frac{m_{\tilde{g}}}{\TEV} \right)^{1/2} \left( \frac{\Lambda_{\rm fid}}{10^7 \ \GEV} \right)^{3/2}\left( \frac{10^{16} \ \GEV}{f_a} \right),
\end{equation}
where we include the factor of $4\pi$ expected from the naive dimensional analysis~\cite{Manohar:1983md,Georgi:1986kr,Luty:1997fk,Cohen:1997rt}.
The mass $m_{\tilde{g}}$ refers to the RG invariant quantity, $m_{\tilde{g},\text{phys}}/g^2$.
This suppression can be understood by the $R$ symmetry in the limit of a vanishing gluino mass, where only a linear combination of an $R$-axion and the QCD axion, which is dominantly the $R$ axion, obtains a mass from the color anomaly. The axion mass is given by $\Lambda_{\rm fid}$ rather than the physical dynamical scale. This can be understood by computing the axion mass in the parameter space where $\Lambda_{\rm fid}= \Lambda_\text{QCD}' $, and extending it to the case with $\Lambda_{\rm fid} < \Lambda_\text{QCD}' $ by holomorphy of the gauge coupling. The MSSM with a large Higgs VEV alone is insufficient in raising the axion mass above the scale of $H_I$ required in Eq.~(\ref{eq:HImin}) to drive the Higgs towards the $D$-flat direction.
Even if we avoid the bound in Eq.~(\ref{eq:HImin}) by small $m_{\rm SUSY}$ in the early universe, the Hubble scale during inflation must be below the MeV scale, which may require fine-tuning in the inflation model parameters.

One can further enhance $\Lambda_\text{fid}$ by introducing additional particles. One possibility involves a moduli field whose VEV controls and increases the gauge coupling constant \cite{Dvali:1995ce,Banks:1996ea,Choi:1996fs,Ipek:2018lhm} during inflation. To satisfy the second criterion of DAMP$_0$, the coupling between the moduli field and $G\tilde{G}$ should be suppressed. The field value of the CP-odd part of the moduli field should remain the same during inflation and in the vacuum.

Another possibility is to extend the MSSM by $N_\Psi$ additional fermion pairs in the $\mathbf{5}+\mathbf{\bar{5}}$ representation of $SU(5)$ as discussed in Ref.~\cite{Jeong:2013xta}
\begin{equation}
\label{eqn:LQCD}
\Lambda_\text{fid} \simeq 10^{7} \ \text{GeV} \left(\frac{\phi_i}{10^{16}~{\rm GeV}}\right)^{2/3} \left(\frac{M_{\Psi,I}}{M_\Psi}\right)^{N_\Psi/9},
\end{equation}
where $M_\Psi$ and $M_{\Psi,I}$ are the vacuum mass and the enhanced mass during inflation, respectively.
We can enhance the mass of $\Psi$ by the large VEVs of some fields, which can be identified with the moduli field discussed above. The minimal example is a coupling with the Higgs, $W\sim H_uH_d \Psi \bar{\Psi}/M$.
For instance, with $N_\Psi = 4$, $M_\Psi = m_{\tilde{g}} = \TEV$, $M_{\Psi,I} = \phi_i = M_{\rm GUT}$, the dynamical scale is $\Lambda_\text{fid} \simeq 10^{13}$ GeV;
the axion mass $m_a \simeq 10^4 \ \GEV (10^{16} \ \GEV/f_a)$ sets an upper bound on $H_I$ since efficient early relaxation to the minimum demands $m_a > H_I$. This is now consistent with Eq.~(\ref{eq:HImin}). To satisfy the second requirement of DAMP$_0$, CP-violating phases of $M_\Psi$ and $M_{\Psi,I}$ should be absent.

Even if  the dynamical scale is raised by the dynamics of fields other than the Higgs, the large Higgs field value during inflation is still crucial. If the Higgs field value is small, the axion mass is suppressed by the small MSSM quark masses as shown in Ref.~\cite{Choi:1996fs}.

We stress that the results of DAMP$_0$ are independent of the specific mechanism that raises the QCD scale during inflation. 
Nonetheless, there are some consistency conditions to be satisfied. The suppression by light quark masses in Eq.~(\ref{eqn:L_fid}) implies that, even when both $\Lambda_\text{QCD}'$ and $\phi_i$ are saturated to the cutoff scale $M$, the Standard Model quark Yukawa couplings $y_q$ set an absolute maximum of $\Lambda_{\rm fid}$
\begin{equation}
\label{eqn:L_fid_max}
\Lambda_{\rm fid}^{\rm max} = M  \prod_{q \in {\rm SM}} y_q^{1/9} \simeq 10^{-2} M.
\end{equation}
Moreover, strong dynamics generates the following effective superpotential for the Higgs,
\begin{equation}
W \simeq \frac{1}{16\pi^2}\left( \left. \Lambda_{\rm fid} \right|_{\phi_i=M} \right)^3  \left( \frac{\phi}{M} \right)^2
\end{equation}
and gives a mass to $\phi$ from the $F$-term potential
\begin{equation}
\label{eq:mhiggs_dyn}
m_\phi \simeq \frac{\left( \left. \Lambda_{\rm fid} \right|_{\phi_i=M} \right)^3}{ 16\pi^2 M^2} ,
\end{equation}
which can dominate and prevent the Higgs from acquiring a large VEV. We require that this should be smaller than $\sqrt{c} H_I$.
Similarly, the scalar superpartner of the axion, the saxion, also receives a mass $m_s \simeq \Lambda_{\rm fid}^3/(f_a^216\pi^2)$. If the saxion mass other than this contribution is also only as large as $\sqrt{c} H_I$, we obtain a stronger constraint when $f_a<M$. The constraint is however absent if the saxion is more strongly stabilized.

\begin{figure}
	\includegraphics[width=0.495\linewidth]{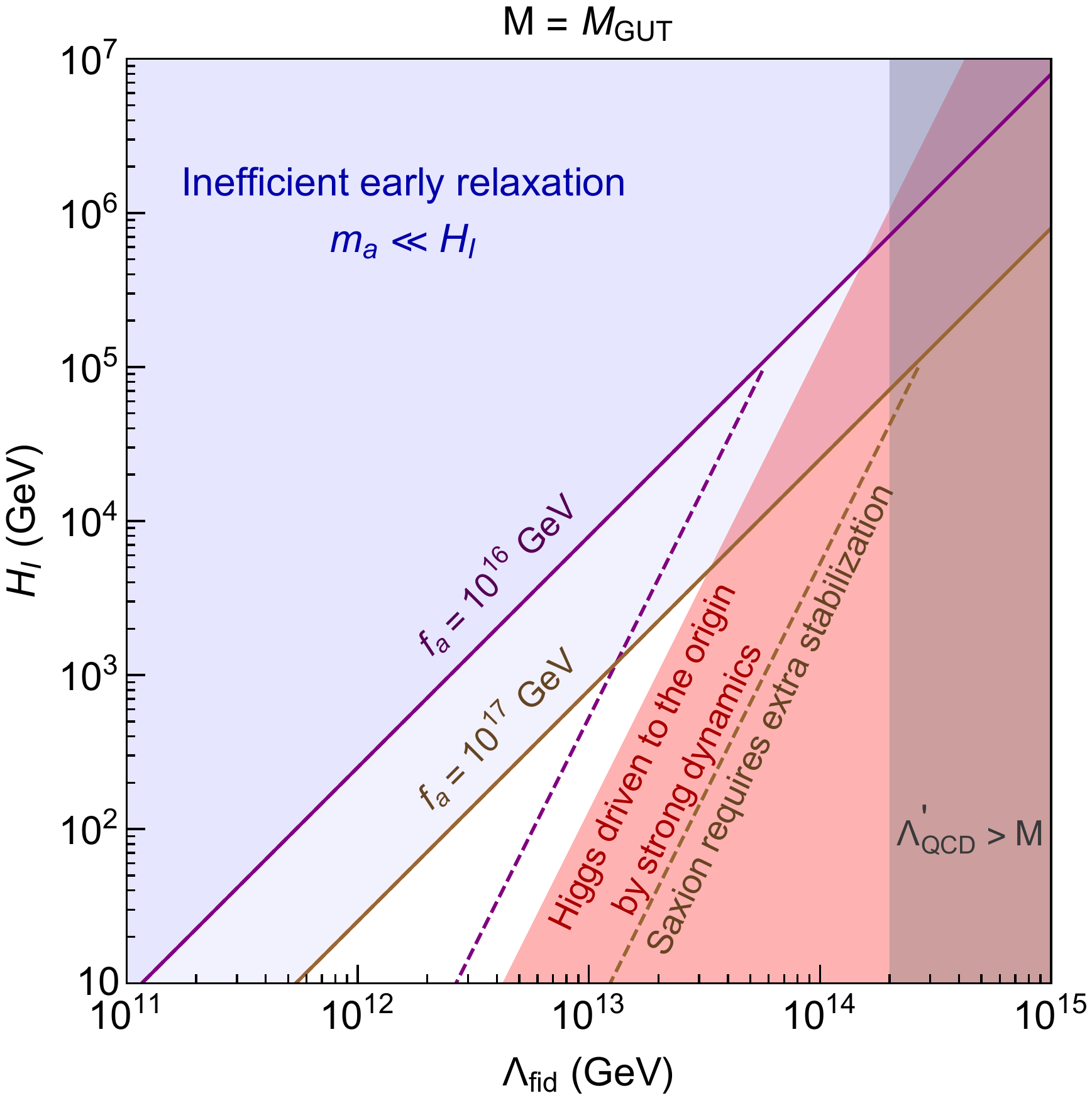}	\includegraphics[width=0.495\linewidth]{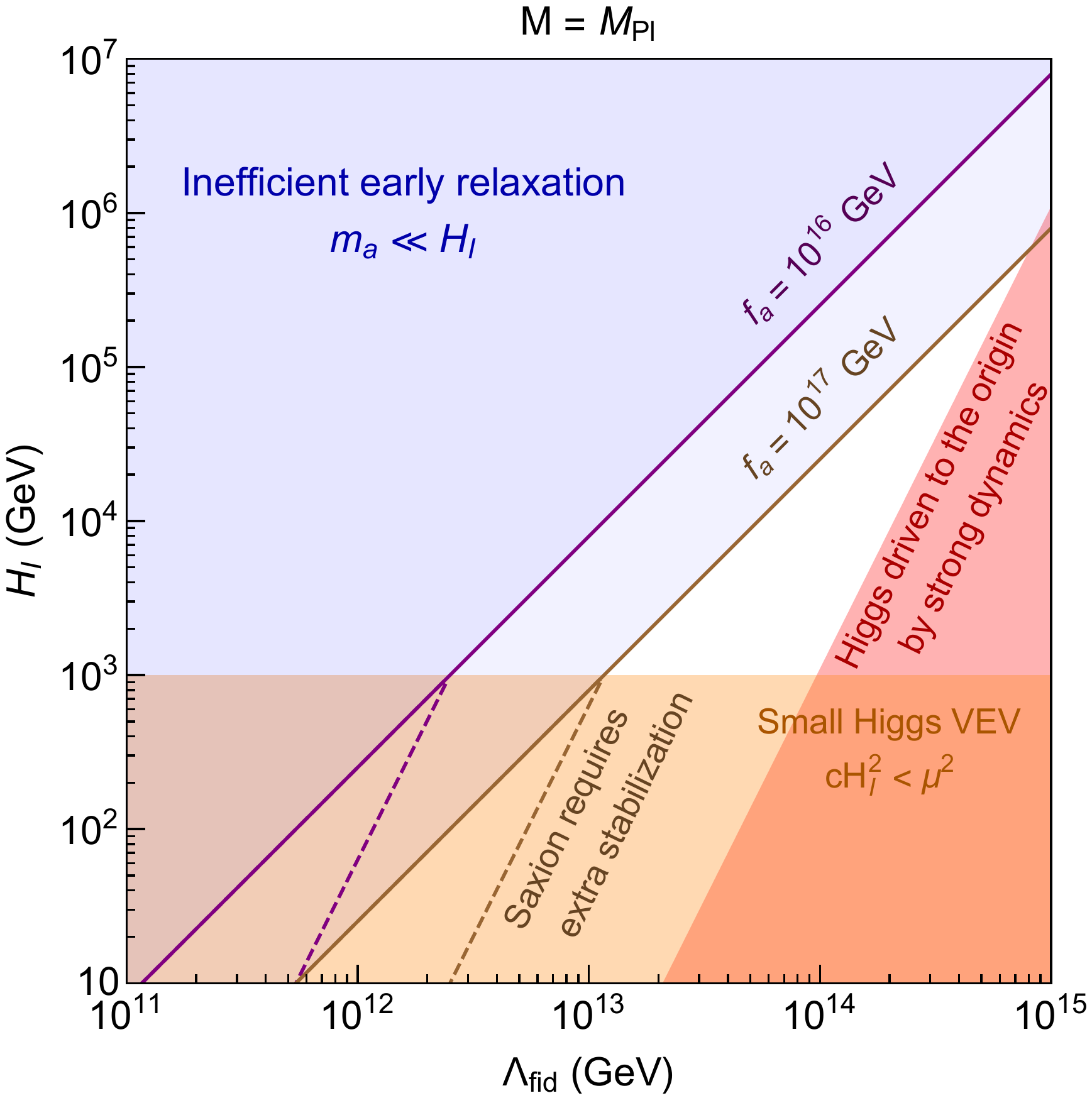}
	\caption{Parameter space for the inflationary Hubble scale $H_I$ and the fiducial confinement scale $\Lambda_{\rm fid}$ defined in Eq.~(\ref{eqn:L_fid}) given $m_{\tilde{g}} = m_\text{SUSY} = \TEV$, and $\phi_i = M$. The left (right) panel is for the cutoff scale $M = M_{\rm GUT} \ (M_{Pl})$ respectively.}
	\label{fig:HIvsLambdaI}
\end{figure}

The constraint on the inflationary Hubble scale $H_I$ for a given fiducial dynamical scale $\Lambda_{\rm fid}$ is shown in Fig.~\ref{fig:HIvsLambdaI}. The blue region reflects the conventional cosmology without DAMP$_0$ for various values of $f_a$ because the axion mass during inflation is less than $H_I$ and the axion field is overdamped by Hubble friction. The gray region is theoretically inaccessible since $\Lambda_{\rm fid}$ exceeds the maximum in Eq.~(\ref{eqn:L_fid_max}). The red region also cannot achieve DAMP$_0$ because strong dynamics generates a Higgs mass in Eq.~(\ref{eq:mhiggs_dyn}) that dominates the Hubble induced mass and drives the Higgs toward the origin.
In the orange region, the Hubble induced mass is subdominant to the SUSY scale in the MSSM and becomes irrelevant for $m_{\rm SUSY} =$ TeV, precluding DAMP$_0$. Lastly, below the dashed line for each labeled value of $f_a$, the saxion mass given by strong dynamics exceeds $\sqrt{c} H_I$, and extra stabilization of the saxion is needed. 
For example, for a chiral multiplet $S$ which non-linearly realizes the PQ symmetry by $S \rightarrow S + iC$, we may add a superpotential~\cite{Harigaya:2017dgd}
\begin{equation}
W = m_S f_a Z e^{-S/f_a} + m_S f_a \bar{Z} e^{S/f_a},
\end{equation}
where $Z$ and $\bar{Z}$ are PQ charged chiral fields. This gives the saxion a mass as large as $m_S$. Since the $F$ terms of $Z$ and $\bar{Z}$ break the supersymmetry, the mass is bounded from above,
\begin{align}
m_S \lesssim m_{3/2} \frac{\mpl}{f_a} \simeq 10^6~{\rm GeV} \frac{10^{16}~{\rm GeV}}{f_a} \frac{m_{3/2}}{10^4~{\rm GeV}},
\end{align}
where $m_{3/2}$ is the gravitino mass. The mass can be large enough for a realistic range of the gravitino mass.

In Fig.~\ref{fig:HIvsLambdaI}, we assume for minimality that the gluino mass during inflation is the same as the vacuum value that we take as $m_{\tilde{g}} = \TEV$. This assumption can be relaxed as well to raise the axion mass further and broaden the allowed parameter space. If $m_{\tilde{g}}$ is larger during inflation, the upper bound on $H_I$ can be raised by a factor as large as $(\Lambda_{\rm fid} / m_{\tilde{g}})^{1/2} = 10^5 \ (\Lambda_{\rm fid} / 10^{13}~\GEV)^{1/2} (\TEV/ m_{\tilde{g}})^{1/2}$ according to Eq.~(\ref{eqn:ma}). This factor is significant and DAMP$_0$ becomes compatible with high scale inflation.

\section{Conclusion}
\label{sec:conclusion}
A large decay constant is not only within the reach of the projected experimental sensitivity for ultralight axions but well-motivated from the theoretical standpoints. Specifically, $f_a \simeq 10^{16}$~GeV can be associated with grand unification, whereas string theory predicts $f_a \simeq 10^{16\mathchar`-17}$~GeV. The QCD axion with such a large $f_a$ faces serious challenges in cosmology, including overproduction of axion dark matter from the misalignment mechanism. It is widely regarded that the misalignment angle $\theta_{\rm mis}$ can be fine-tuned to avoid this issue. Even in this case, however, isocurvature perturbations are in conflict with high scale inflation.

In this paper, we identify a class of models where the misalignment angle is set to a small value due to axion dynamics instead of mere fine-tuning. The conventional assumption of the axion field during inflation is such that Hubble friction dominates over the axion mass and the field value remains constant. Nevertheless, there are various scenarios where the axion mass can be much larger so that the axion is relaxed to the minimum in the early universe. We refer to such damping mechanism as Dynamical Axion Misalignment Production (DAMP). Furthermore, if the model possesses an approximate CP symmetry, the minima in the early universe can nearly align with that of today, leading to $\theta_{\rm mis} \simeq 0$, which we call DAMP$_0$. This early relaxation to a minimum close to today's value resolves both the axion overproduction and isocurvature difficulties.

We realize the DAMP$_0$ scenario using supersymmetric models where the coupling between the Higgs and the inflaton results in a large axion mass during inflation. In particular, the inflaton energy density can induce a negative mass term for the Higgs, resulting in large Higgs VEVs in the $D$-flat direction. The quark masses that become larger due to the Higgs VEVs modify the RG running of the strong coupling constant and bring about a much larger QCD confinement scale. The axion mass is then enhanced by the high QCD scale, fulfilling the criteria for DAMP$_0$ in a consistent cosmology. We can also raise the QCD scale by means of generic moduli fields. The large Higgs field value is still crucial since otherwise the axion mass is suppressed by the small MSSM quark masses. In summary, the axion misalignment near the bottom of the potential can result from early dynamics and hence a small $\theta_{\rm mis}$ can be regarded as a prediction of the model rather than fine-tuning.

 {\bf Acknowledgment.}---%
The authors thank Lawrence J. Hall and Andrew J. Long for discussions. K.H. thanks the Leinweber Center for Theoretical Physics for the warm hospitality he received during his visit when most of this work was done. The work was supported in part by the DoE Early Career Grant DE-SC0019225 (R.C.) and the DoE grant DE-SC0009988 (K.H.).

\bibliography{bibtexrefs}

\end{document}